\documentclass[a4paper,11pt]{article}
\usepackage{pos}
\usepackage{color}
\usepackage{url}
\usepackage{lineno}
\usepackage{sidecap}
\usepackage{textcomp}
\usepackage{titlesec}
\usepackage{setspace}
\titlespacing*{\section} {0pt}{3.0ex plus 1ex minus .2ex}{1.5ex plus .2ex}
\titlespacing*{\subsection}{0pt}{3.0ex plus 1ex minus .2ex}{1.5ex plus .2ex}

\title{IceCube's response to supernovae and periodic features in the count rates}
 \ShortTitle{MeV Neutrinos from Supernovae in IceCube}
\author{The IceCube Collaboration \\{\normalsize \normalfont(a complete list of authors can be found at the end of the proceedings)}}
\emailAdd{afritz04@uni-mainz.de}
\emailAdd{dkappesser@uni-mainz.de}

\abstract{
The IceCube Neutrino Observatory is highly sensitive to neutrino bursts of $\mathcal{O}$(10) MeV energy that are would be generated by core collapse supernovae in our Galaxy. It will resolve temporal structures in supernova light curves particularly well. 
In the light of an improved understanding of the ice properties and the detector response, the effective area and the corresponding uncertainties were newly determined with a Geant4-based Monte Carlo. Uncertainties due to cross sections and oscillation effects in the Earth were also investigated.  This analysis has been extended by simulating a very large sample to determine the small coincidence probability between optical modules that bears information on the average neutrino energy. 
These simulation results were then used to interpret the data in time and frequency space. 
While the availability to record data for low energy neutrinos from 
supernovae is close to perfect (99.2\% between 2013-2020), the analysis requires that the detector works faultlessly and artifacts do not mimic the signal in the 13 years of data taken so far. An effort has been made to keep the uptime after all analysis steps similarly high. The frequency space can be studied in a range between 1 Hz and 1/year to test the detector stability with high accuracy, to study the influence of cosmic rays, and to search for periodic phenomena that lead to sub-threshold increases in the count rates.
Here we discuss the results of the simulations and the corresponding systematic limitations, the method to reconstruct the mean neutrino energy for a recorded supernova, as well as aspects of the analyses of continuously taken optical module rate data in the time and frequency domain.

\vspace{4mm}
{\bfseries Corresponding authors:}
Alexander Fritz$^{1*}$, David Kappesser$^{1}$\\
{$^{1}$ \itshape Johannes Gutenberg Universit\"at Mainz,\\Staudingerweg 9, D-55270 Mainz, Germany} 
\\[4mm]
$^*$ Presenter

\FullConference{37$^{\rm{th}}$ International Cosmic Ray Conference (ICRC 2021)\\
		July 12th -- 23rd, 2021\\
		Online -- Berlin, Germany}

}
\begin{document}
\maketitle

\section{Introduction}
IceCube, which includes the more densely instrumented DeepCore subarray, is an excellent detector for measuring the neutrino 
light curve of a galactic supernova with high temporal precision. 
Neutrinos from a galactic supernova would interact in the deep clear glacial ice at the South Pole and add hits to the low $\mathcal{O}$(0.5) kHz dark rate of IceCube’s 5160 optical modules. While individual  neutrinos would not trigger the IceCube detector due to their low energy, the collective rise in count rates during an $\mathcal{O}$(10) s time span can be statistically separated to provide a trigger. 
The physics capabilities of IceCube for supernova detection were summarized 
in~\cite{bib:AA} and improvements were discussed in ICRC contributions~\cite{bib:icrc2019} and other conferences~\cite{bib:koepke}. 

In the light of improvements of the understanding of detector and ice properties, we undertook a detailed Geant4-based study of the effective volume for supernova neutrino detection, updated the simulation with state-of-the-art cross section determinations and implemented a recent improved description of single photon-induced PMT pulses~\cite{bib:spe}. 

We concentrate on the dominant  $\bar{\nu}_\mathrm{e} + \mathrm{p} \rightarrow \mathrm{n} + \mathrm{e^+}$ reaction (inverse beta decay) with subsequent capture of the neutron. The positron emits $\approx 180$ Cherenkov photons per MeV. Compton scattered electrons from 2.2 MeV neutron capture photons add roughly 100 photons on average. For $E_\nu\lessapprox$ 20 MeV, reactions with oxygen are negligible. 

Due to the large distance between IceCube's optical sensors (DOMs), individual MeV neutrino interactions cannot be reconstructed. 
Km3NeT~\cite{bib:km3net} 
relies on coincidences between the sensors of  multiple PMT optical modules to improve their supernova detection. The method lowers the effective dark rate, extends the detection horizon, and allows one to determine the average energy of the detected neutrinos. While  multiple PMT optical modules will be used in the ongoing IceCube upgrade and in the future IceCube-Gen2 detector for similar studies~\cite{bib:muenster}, the IceCube DOM houses only 
one 10" PMT. Due to the low single photon resolution of PMTs, the average energy of detected neutrinos needs to be determined from coincidences between different optical modules that are $17\,$ m (IceCube) or $6\,$ m and 10\, m (DeepCore) apart vertically and 40 -- 125 m apart horizontally. The probability for such coincidences, which scales roughly quadratic with the neutrino energy, is very small ($\approx$ 0.04 (0.15) percent in IceCube (DeepCore) at $E_{\nu}=25$ MeV). Using such an approach, earlier simplified simulations confirmed that one may indeed obtain a meaningful measure of the average $\bar{\nu}_\mathrm{e}$  energy~\cite{bib:salathe,bib:bruijn}. Here we present a more rigorous determination using IceCube's simulation and reconstruction tools.

Such simulations are needed for interpreting the results of ongoing searches for supernovae and periodic countrate features in 13 years of IceCube data, for which we will present selected aspects. 

\section{Simulation} 
The signal hit rate $r_\mathrm{SN}$ per DOM for the inverse beta decay is  given by
\begin{eqnarray}
	r_\mathrm{SN}(t) & \approx & \frac{n_\mathrm{target} \,L_{\bar{\nu}_\mathrm{e}}}{4\pi d^2} \; \frac{V_\mathrm{e^+}^\mathrm{eff}}{\overline{E}_{\bar{\nu}_\mathrm{e}}(t)}\;
	 \int_0^{\infty} \,dE_\mathrm{e^+} \!\! \int_0^{\infty}  \,dE_{\bar{\nu}_\mathrm{e}} 
    \times\,
	\frac{d\sigma}{dE_\mathrm{e^+}}(E_\mathrm{e^+},E_{\bar{\nu}_\mathrm{e}}) \,  \,f (E_{\bar{\nu}_\mathrm{e}},\overline{E}_{\bar{\nu}_\mathrm{e}} , \alpha,t) \, ,  \label{eq:rate}
\end{eqnarray}
where $f(E_{\bar{\nu}_\mathrm{e}},\overline{E}_{\bar{\nu}_\mathrm{e}},t)$ is the time dependent probability density function describing the neutrino energy distribution with average $\overline{E}_{\bar{\nu}_\mathrm{e}}=\int_0^{\infty} dE_{\bar{\nu}_\mathrm{e}}\; E_{\bar{\nu}_\mathrm{e}} \;f(E_{\bar{\nu}_\mathrm{e}},\overline{E}_{\bar{\nu}_\mathrm{e}} , \alpha,t)$.
$L_{\bar{\nu}_\mathrm{e}}$ denotes the supernova energy luminosity, 
$n_\mathrm{target}$ is the density of targets in the ice, $d$ is the distance to the supernova. 
The effective volume $V_\mathrm{e^+}^\mathrm{eff}$ for a positron  subsumes all media, detector and analysis effects. 
 
Very roughly speaking,   $V_\mathrm{e^+}^\mathrm{eff}/\overline{E}_{\bar{\nu}_\mathrm{e}}$ may be estimated
by the product of absorption length, Cherenkov photons per unit energy, DOM geometric
cross section, optical
module sensitivity, and the fraction of single
photon hits that pass the electronic threshold for a single photon~\cite{bib:oldpapers,bib:AA}.
Using approximations, one can analytically estimate~\cite{bib:icrc2019} the rate of hits in IceCube from the inverse beta decay interaction for a given model with $\approx 25\%$ precision.
To go beyond such an analytic estimate, the response to supernova neutrinos interacting in the IceCube detector can be simulated with programs of increasing detail and demand on computing resources:
\begin{itemize}
    \item SNOwGLoBES~\cite{bib:snowglobes}: a fast and simple mean event rate calculator for low-energy neutrino interactions using the analytic formula discussed above.
    \item Asteria~\cite{bib:asteria}: a parameterized IceCube Monte Carlo.
    \item Full Geant4-based Monte Carlo using IceCube software tools (this paper). 
\end{itemize}
\subsection{Geant4-based IceCube Monte Carlo}
\begin{SCfigure}
\vspace{0.5cm}
\includegraphics[angle=0,width=0.59\textwidth]{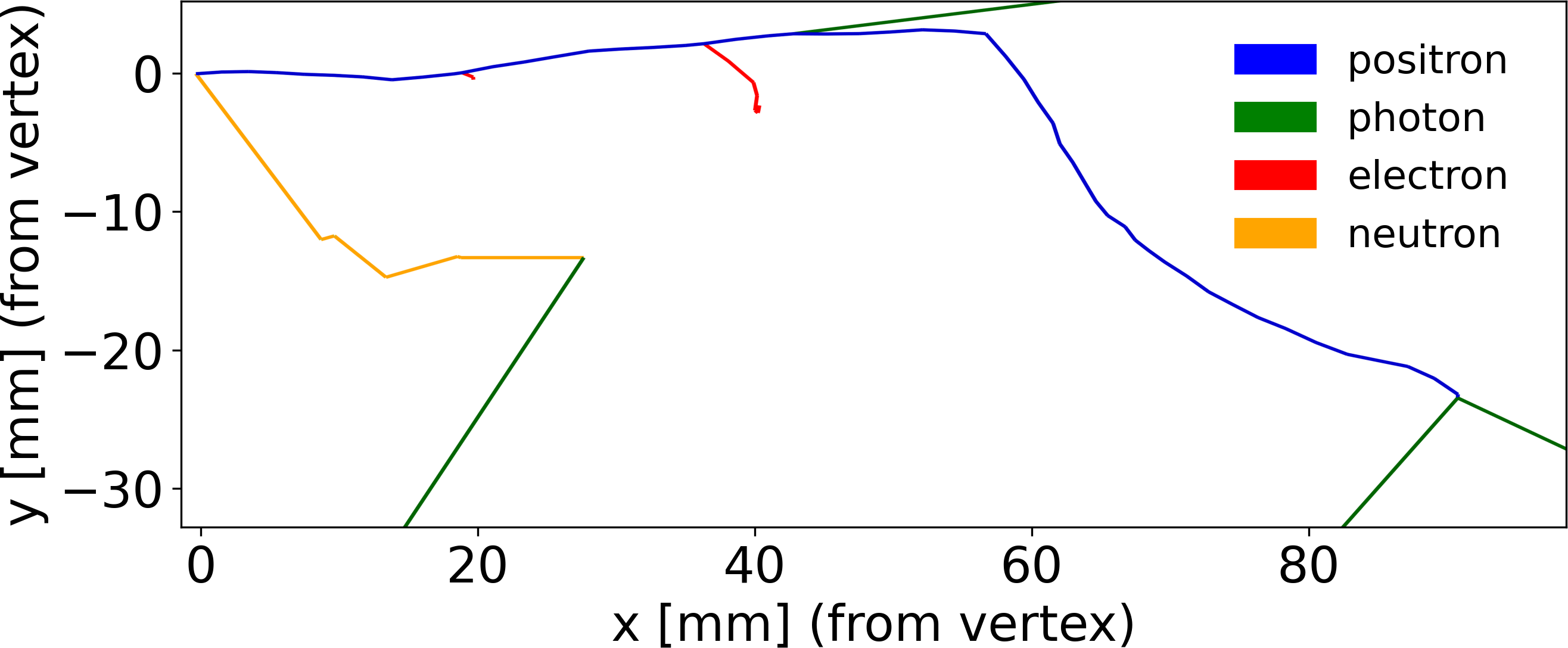}
\vspace*{-0.8cm}
\caption{\small  Geant4 simulated $\bar{\nu}_\mathrm{e}+ \mathrm{p} \rightarrow \mathrm{n} + \mathrm{e^+}$ event projected to x,y plane: 15 MeV $\mathrm{e}^+$ (blue) entering from left emitting a bremsstrahlung $\gamma$ 
(green) and two $\gamma$'a from annihilation. The neutron (orange) 
is captured with a subsequent emission of a 2.24 MeV $\gamma$. Low energy $\mathrm{e}^-$ from ionization are shown in red. }
\label{fig:geant}
\end{SCfigure}

We generate positrons from the inverse beta decay with directions and energies corresponding to the properties of an incident neutrino plus delayed neutron capture photons. A typical interaction of a 15 MeV positron in ice is shown in Fig.~\ref{fig:geant}. Further steps are performed using standard IceCube Software, Geant4 to simulate particle interactions, and a custom GPU-optimized Cherenkov photon tracker~\cite{bib:clsim}. Fourteen samples of mono-energetic positrons were generated with 1 billion interactions each and samples about the expected mean neutrino energy were extended to 4 billion. Additional samples with 0.1 billion (1 billion for nominal values) interactions were created to test variations in the modeling of the ice. 
On average, roughly every 500th positron leads to at least one photon detected by IceCube. One billion interactions yield about 250 neighboring DOM coincidences. 90\% and 95\% of these coincidences have their corresponding interaction vertices located within 20 meters of the DOM that receives the first hit in IceCube and DeepCore, respectively. Instead of simulating PMT noise and cosmic ray muons, unbiased IceCube DOM hit rates are overlaid on the Monte Carlo events.   
\subsection{Effective volume and systematic  uncertainties}
\begin{figure}
\includegraphics[angle=0,width=0.57\textwidth]{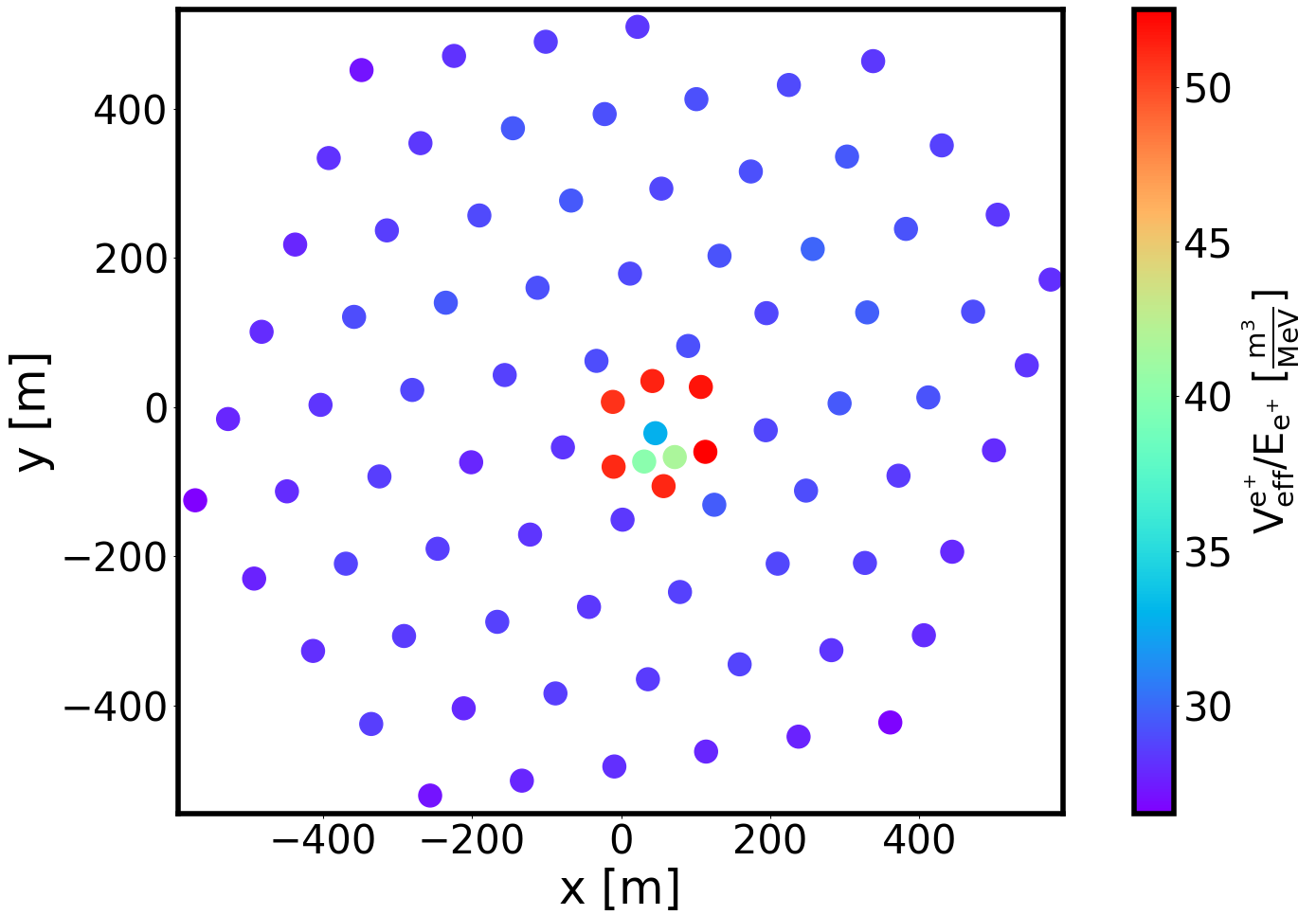}
\includegraphics[,angle=0,width=0.43\textwidth]{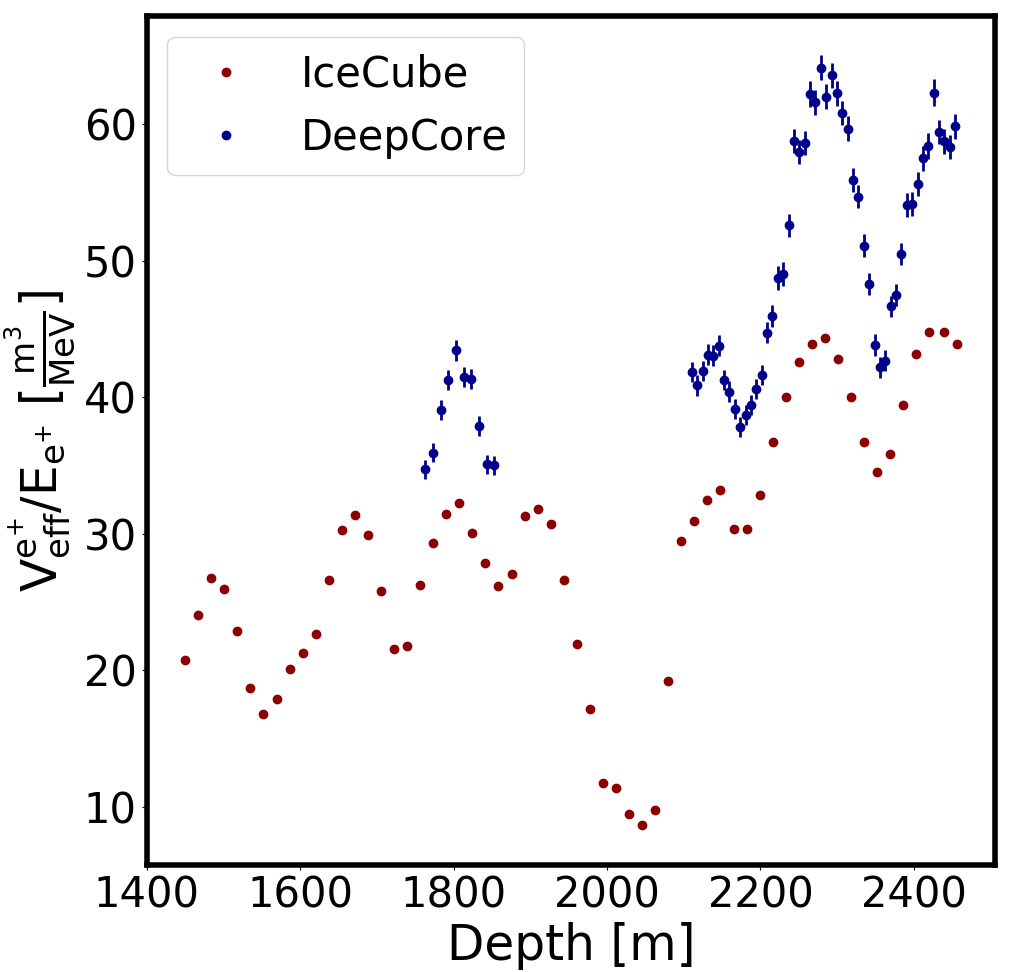}
\caption{\small Left: effective positron volume as function of IceCube string position. Right: effective positron volume as function of depth for IceCube (red) and DeepCore (blue).}
\label{fig:effectivevolume}
\end{figure}

Since inverse beta decays 
dominate, we discuss the effective positron volume per neutrino energy $V_\mathrm{e^+}^\mathrm{eff}/\overline{E}_{\bar{\nu}_\mathrm{e}}$ in units of m$^3/$MeV. 
IceCube is embedded in the Antarctic ice shield with slightly inclined layers of dust and other scattering centers. These impurities, as well as the pressure dependent optical properties of air-loaded ice, lead to depth-dependent absorption and scattering lengths. The most prominent dust layer is situated around 1950 m depth; below, the ice is much clearer. 
Figs.~\ref{fig:effectivevolume} (left) shows a bird eye view of the IceCube observatory with the color code indicating $V_\mathrm{e^+}^\mathrm{eff}/\overline{E}_{\bar{\nu}_\mathrm{e}}$. The strings with large effective volumes belong to DeepCore, which  is mostly embedded in the clear ice region below a central dust layer. This becomes evident from the right hand plot, where $V_\mathrm{e^+}^\mathrm{eff}/\overline{E}_{\bar{\nu}_\mathrm{e}}$ is plotted as a function of depth for IceCube DOMs (red) and DeepCore DOMs (blue). In addition to populating clear ice, the 360 DeepCore DOMs house photomultipliers with a quantum efficiency that is about 1.35 larger than that of standard IceCube DOMs. 

\begin{figure}
\includegraphics[angle=0,width=0.47\textwidth]{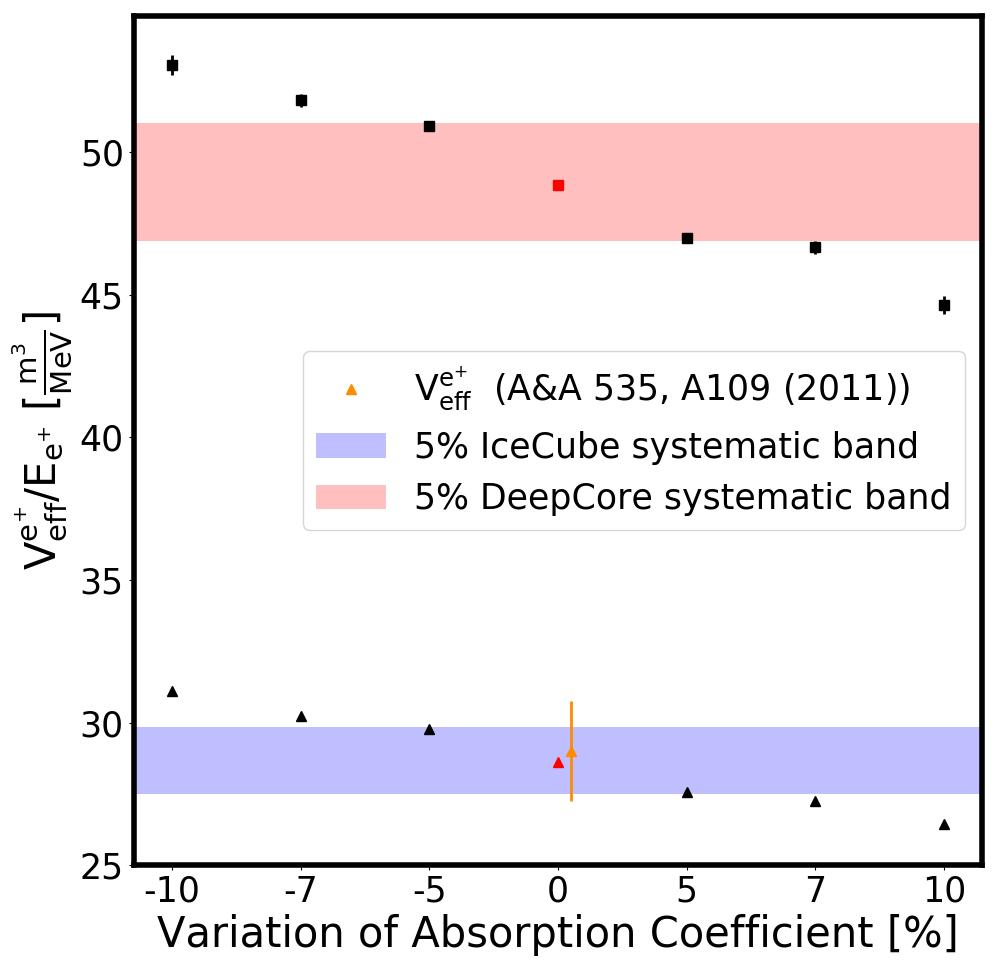}
\includegraphics[angle=0,width=0.52\textwidth]{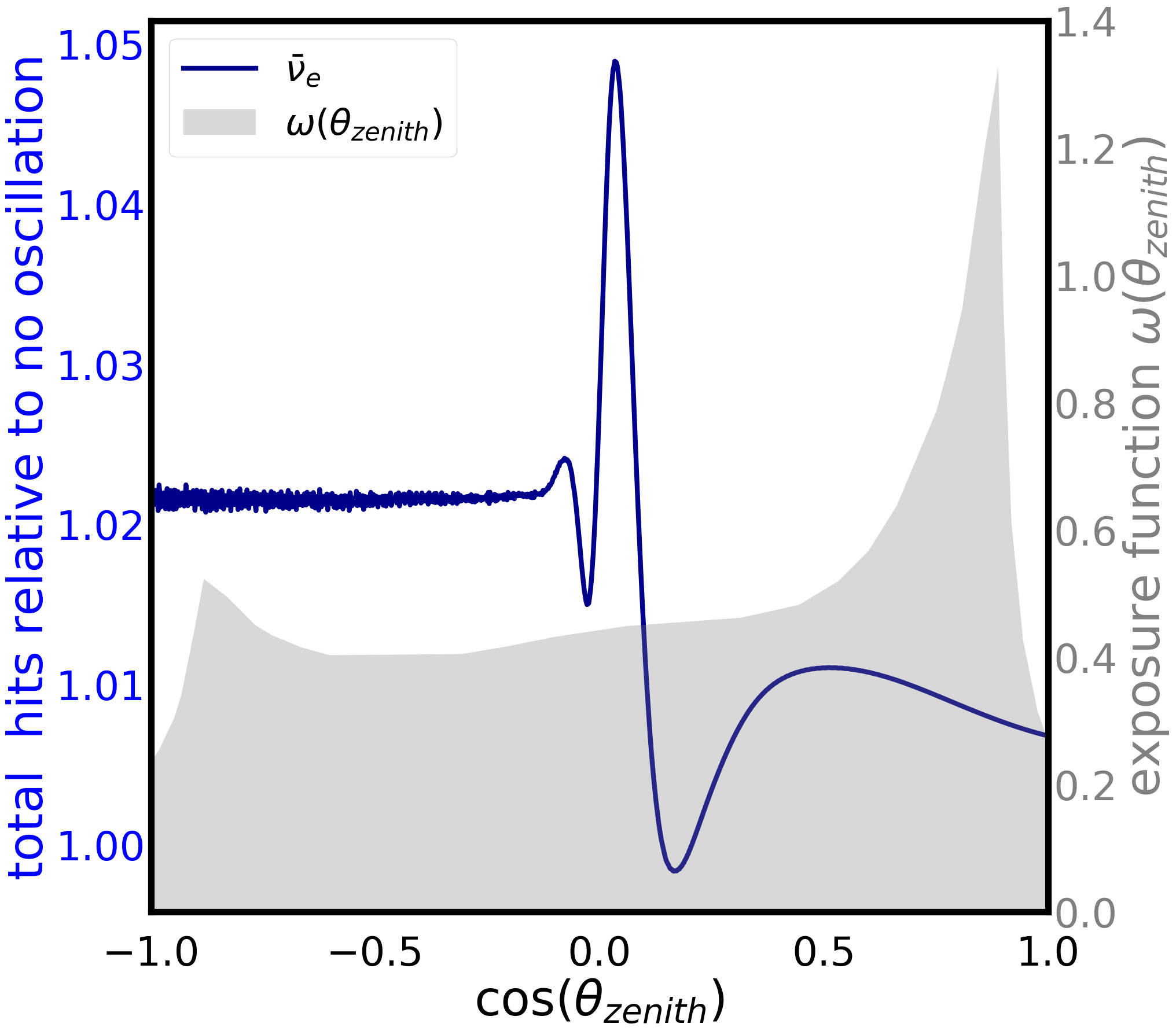}
\caption{\small Left: Systematic uncertainties due to absorption in the ice for IceCube (blue) and DeepCore (red). 
The dependence on the scattering is very small (see text). 
For comparison, the published value (in orange) is shown including its uncertainty. The upper and lower bands correspond to  5\% uncertainties in the scattering and absorption coefficients, respectively. Right: Energy weighted ratio of registered hits with and without Earth oscillations, integrated over the full time window  for model~\cite{bib:huedepohl}.
Also shown is the exposure from potential supernovae in the galactic plane (from~\cite{bib:mirizzi}, Fig.3, solid curve). 
The highest probability is in the South ($\cos\theta\rightarrow 1$).}
\label{fig:systematics}
\end{figure}

While a remarkable effort has gone into the in-situ calibration of the ice properties, there are  uncertainties associated with it~\cite{bib:ICRCice}. The ice density is known to better than 0.2\%. However, the uncertainties on the scattering and absorption coefficients are presently estimated at 5\% each. Fig.~\ref{fig:systematics} (left) shows the result of studies with Monte Carlo samples, where the scattering and absorption lengths were varied within 10\%. A quantitative evaluation shows  correlation between the relative uncertainties of effective volumes and absorption coefficients of
$-0.78 \pm 0.02$ for IceCube and 
$-0.81 \pm 0.04$ for DeepCore. The correlation with the scattering length, on the other hand, is very small: $0.037 \pm 0.015$ and $-0.018\pm 0.037$ for IceCube and DeepCore, respectively.   The color bands in Fig.~\ref{fig:systematics} (left) show the 5\% uncertainty contours for IceCube  and DeepCore; the result published in ~\cite{bib:AA} is  compatible with the new determination that uses the best current knowledge of the ice properties. 
Ice properties are not the only source of detector-related uncertainties (see Table~\ref{tab:systematics}). For example, the absolute DOM efficiency in-situ is presently known to 10\%.  
In addition, there are uncertainties on the cross sections. Neutrino interactions with oxygen are poorly known; however, they only play a role at neutrino energies beyond 20 MeV (see Tab.~\ref{tab:systematics}).
Their contributions for 8.8 $M_\odot$ progenitor~\cite{bib:huedepohl} and Black Hole forming~\cite{bib:bh} models are only 1\% and 14\%, respectively.

We also studied potential uncertainties due to neutrino oscillations in the matter of the Earth. These become relevant when comparing the results of detectors at locations with different neutrino path lengths in the Earth or when the supernova position is unknown. Oscillation effects only play a role if either cross sections, fluxes, or energy spectra vary between flavors. While these are similar during the cooling phase, substantial differences in the early phase of neutrino emission may strongly modify the neutrino light curve. Fig.~\ref{fig:systematics} (right) shows the energy weighted ratio of total registered hits with and without Earth oscillations  as function of the zenith angle for model~\cite{bib:huedepohl}. If studied as function of time (not shown),  the largest effect (factor 2.3) is seen around 0.015 s. Note that the highest probability for a galactic supernova is in the Southern sky, where neutrinos do not cross the Earth's core and the oscillation effects are smooth (gray shaded exposure area in  Fig.~\ref{fig:systematics}, scale on right).
Finally, large uncertainties in the modeling of supernovae may remain even if an optical counterpart can be studied in detail, as well as MSW and collective neutrino oscillations in the core of the developing supernova.  Addressing these goes beyond the scope of this paper. 
\begin{table}[h]
    \centering
    \begin{tabular}{|c|c|}
    \hline
        mean $\mathrm{e}^\pm$ track length in ice & 5\% \\
       positron effective volume & $_{-13}^{+15}$ \% \\
     cross sections (inverse beta, $\mathrm{e}^-$ scattering, oxygen) & <1\%, <1\%, [0.2, 1.4]\% \\
     angle dependent MSW Earth oscillation  & [-0.2, 4.9]\% \\
     \hline
    \end{tabular}
    \caption{Summary of systematic uncertainties on the photon flux from supernovae (detector and media properties, cross sections for models~\cite{bib:huedepohl,bib:bh}, variation in total hits from Earth oscillations for model~\cite{bib:huedepohl}).}
    \label{tab:systematics}
\end{table}
\vspace{-30pt}
\subsection{Average energy of detected neutrinos from coincidences between DOMs}
For supernova neutrino candidate events,
raw detector data are stored that allow one to determine the noise and signal rate with high timing accuracy and to check for coincidences between modules~\cite{bib:hitspool}. We restrict ourselves to coincidences between the nearest neighbors. The average neutrino energy can be determined using 
a Poissonian likelihood with the following parameters:
\begin{eqnarray}
N_\mathrm{comb} & \widehat{=}& \mathrm{number\; of\; combinations}\nonumber\\ 
r_\mathrm{noise},\;r_\mathrm{signal}  & \widehat{=} & \mathrm{noise\; and \; signal\; rates}\nonumber\\
r_\mathrm{coinc} \approx a\cdot r_\mathrm{signal}\cdot \overline{E}_{\mathrm{e}^+}& \widehat{=} & \mathrm{coincidence\; rate\; from\; same\; neutrino\; interaction}\nonumber\\
N(T) & \widehat{=} & \mathrm{number\; of\; observed\; events\; for\; duration\; of\; experiment\;} T \nonumber
\end{eqnarray}
The number of coincidences depends on the gate width $\Delta t$ and thus on the module distance and scattering length. The coefficient $a$ 
must be determined by Monte Carlo. Optimized gate widths $-64 (-96) $ns $ <\Delta_t<128 (192)$ ns  for IceCube (DeepCore) were found~\footnote{$\Delta t < 0$ denotes a first hit in the upper DOM and $\Delta t > 0$ refers to a first hit in the lower DOM.} by comparing the time distributions of noise-related coincidences, which show a flat behavior, coincidences arising from supernova neutrinos and  cosmic ray muons, that result in a shifted peak. 

\begin{figure}
\includegraphics[angle=0,width=0.48\textwidth]{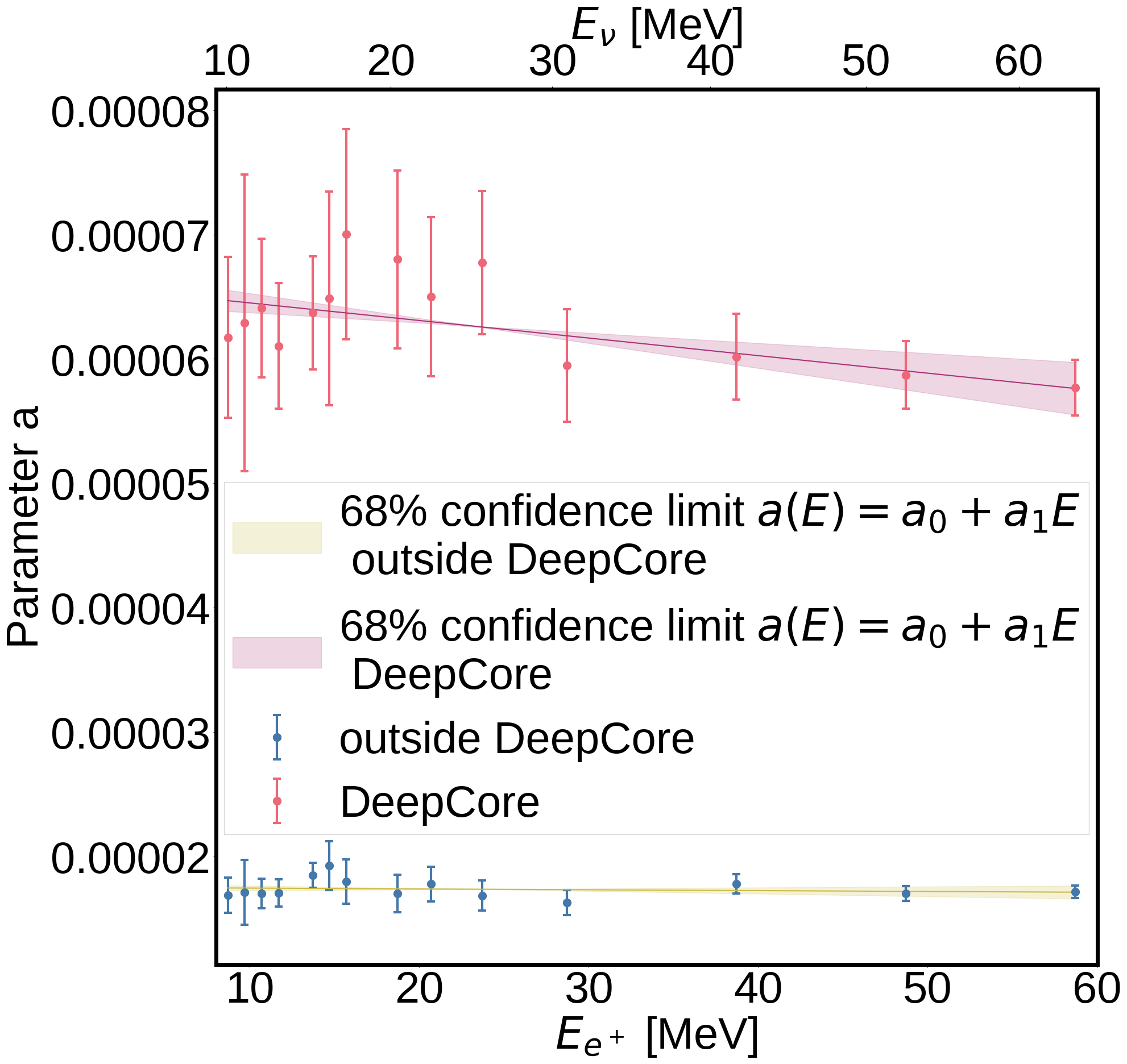}\vspace{-0.3cm}
\includegraphics[angle=0,width=0.51\textwidth]{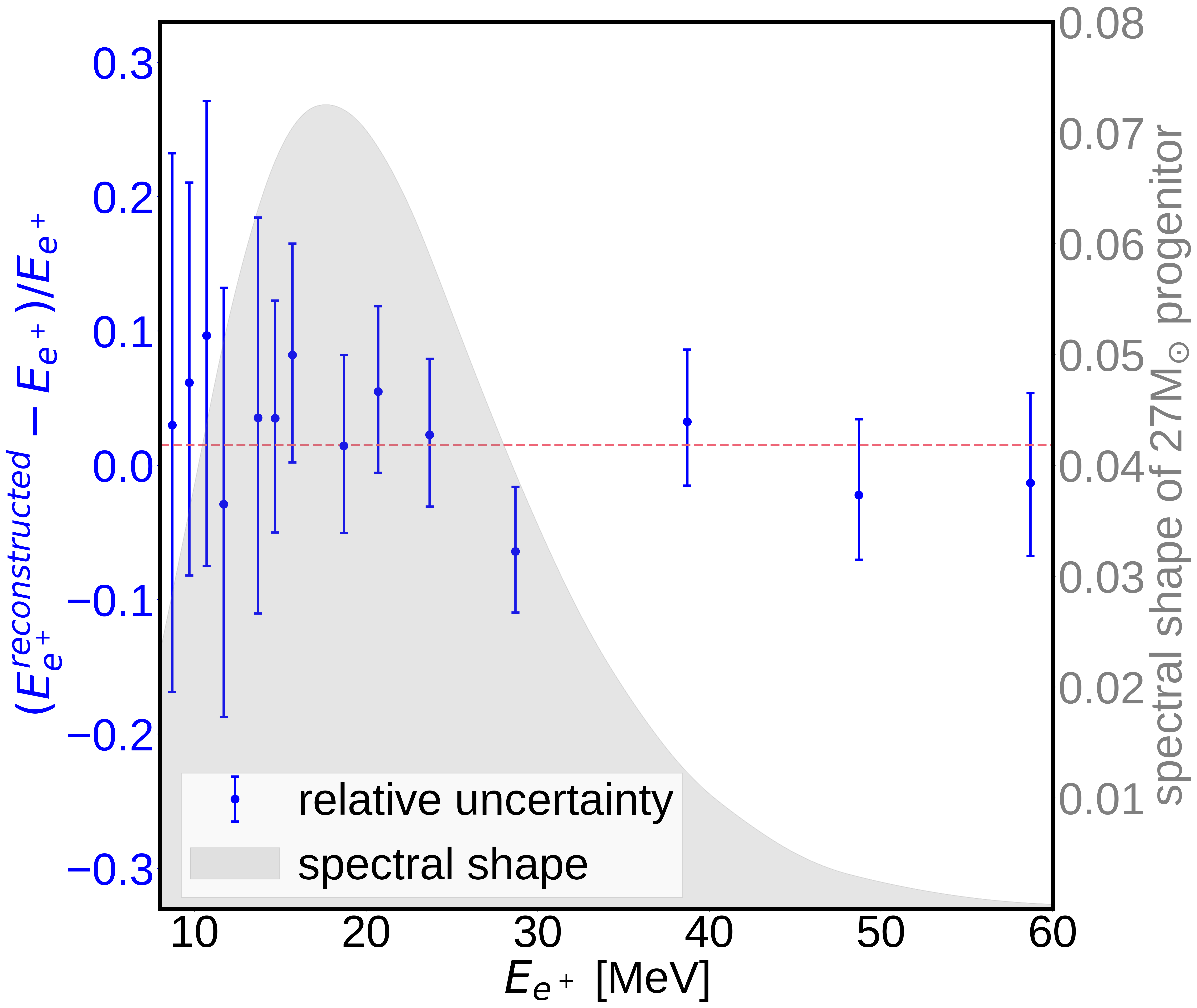}
\caption{\small Left: 
Fits to determine the parameter $a(E_{\mathrm{e}^+})=a_0 + a_1\cdot E_{\mathrm{e}^+}$ (see Eq.~\ref{eq:likel}) with associated 68\% uncertainty belts. The corresponding $E_nu$ scale is shown on top. 
 Right: Relative average positron energy resolutions determined from the likelihood fit. 
Each error bar correspond to 1 billion injected neutrinos.
The reconstructed energy (dashed horizontal line) is slightly underestimated. The probability density of interacting neutrinos for a 27 $m_\odot$ progenitor (gray scale on the right) indicates the region of interest.}
\label{fig:coinc}
\end{figure}

Coincidences occur between noise pulses, noise and signal pulses, pulses from different signal events, and pulses from the {\em same} neutrino interaction.
Disregarding all constants, the combined likelihood is a product of likelihoods with $i\in$ [DeepCore, IceCube]:
\begin{eqnarray}
-\log\mathcal{L}(\overline{E}_{\mathrm{e}^+})&=&\sum^\mathrm{ensemble}_{i}T\left[ N_\mathrm{comb,i}\cdot\Delta t_i \left(\epsilon\cdot r_\mathrm{noise}^2+ \epsilon'\cdot r_\mathrm{noise} r_\mathrm{signal}+ \epsilon''\cdot r_\mathrm{signal}^2+ \frac{r_\mathrm{signal}\cdot a_i\overline{E}_{\mathrm{e}^+}}{\Delta t_i}\right)\right .\nonumber\\
& - &\left . N\cdot \log\left(\epsilon\cdot r_\mathrm{noise}^2+ \epsilon'\cdot r_\mathrm{noise} r_\mathrm{signal}+ \epsilon''\cdot r_\mathrm{signal}^2+ \frac{r_\mathrm{signal}\cdot a_i\overline{E}_{\mathrm{e}^+}}{\Delta t_i}\right)\right]\;\label{eq:likel}.
\end{eqnarray}
 For pure Poissonian noise without contributions from atmospheric muons, $\epsilon=$ \textonehalf$\,\epsilon'=\epsilon''=1$; in practice, $\epsilon,\epsilon'$ need to be determined as nuisance parameters from data and Monte Carlo. 
 For an ideal detector, $a_i=\mathrm{const}$. Fig.~\ref{fig:coinc}  (left) shows a fit to the $a_i$ parameters for each simulated energy, showing a slight energy dependence due to the finite detector volume. The relative energy resolution and bias can be deduced from Fig.~\ref{fig:coinc} (right). Interestingly, the dense DeepCore strings contribute more to the resolution than the 13 times larger number of DOMs in the IceCube strings. The bias (dashed red line) is consistent with zero, the resolution improves as expected with energy. The actual resolution depends on the number of coincident photons and thus on the distance and the model. 
 The resolution shown in Fig.~\ref{fig:coinc} corresponds to a 27 $M_{\odot}$ progenitor model at 4 kpc distance with 24 MeV average energy of the interacting neutrinos ($E_{\mathrm{e}^+}$ spectrum  shown in gray). 
\color{black}

\section{DOM hit rate analyses in time and frequency space}
The simulation results discussed so far will be used to quantify the analysis of 13 years of IceCube DOM hit rates in time and frequency space. After a brief introduction of the rate data, we will concentrate on a method to study the corresponding frequency space. 

DOM rates are continuously counted in 1.6384 ms time bins by the on-board firmware. A dedicated online software system re-bins the data to 2 ms and searches the data stream for collective rate increases that are characteristic of supernovae in various choices of time bins. Data binned in 500 ms intervals are available independent of a trigger.
The data acquisition system incorporates an artificial dead time of 250 $\mu$s to suppress photons from glass luminescence, which -- on average -- lowers the DOM noise rate to about 300 Hz. A contribution from atmospheric muons is corrected for both online and offline by unfolding hits from triggered muon tracks.  

\begin{figure}
\includegraphics[angle=0,width=0.5\textwidth]{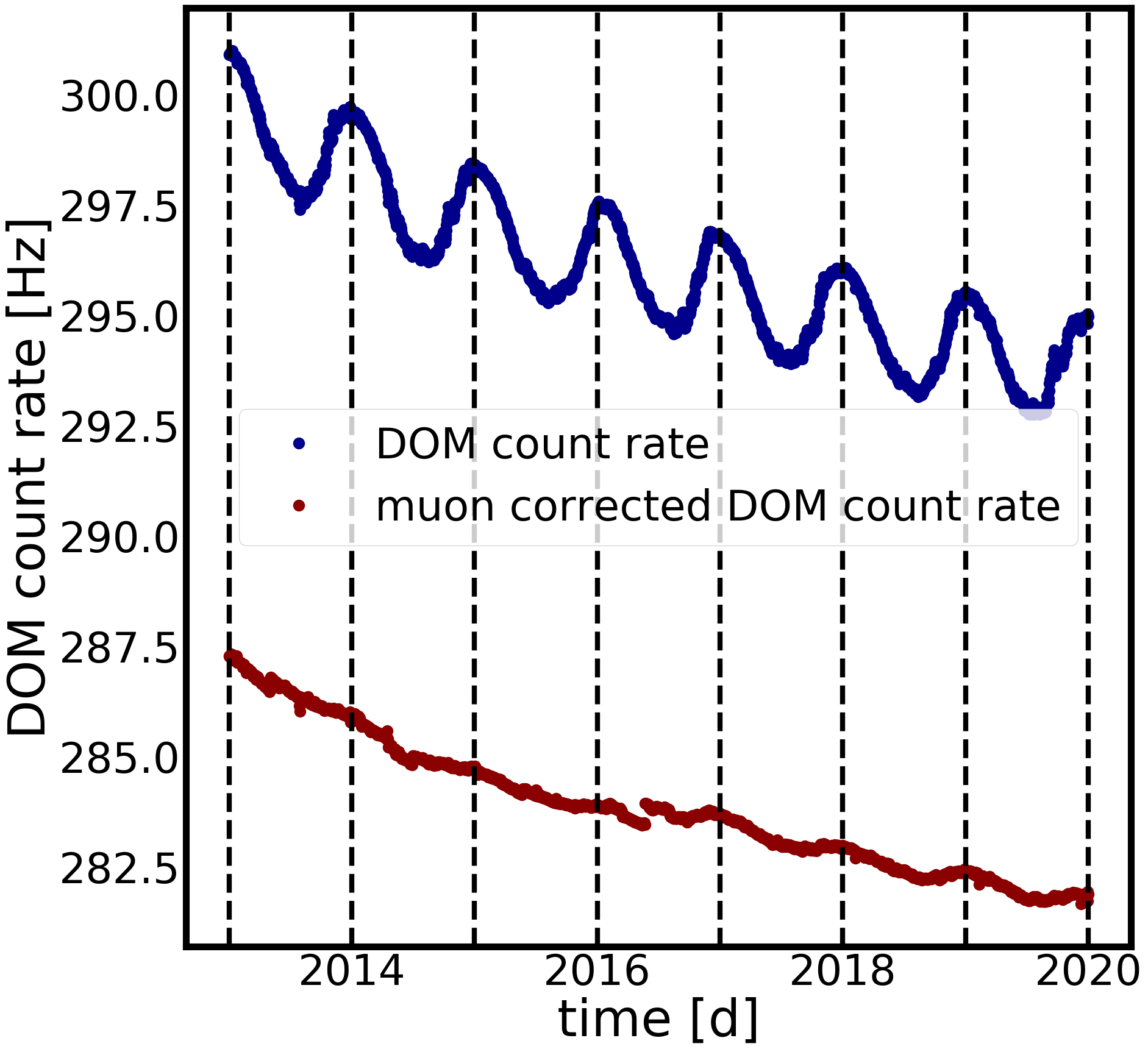}
\includegraphics[angle=0,width=0.475\textwidth]{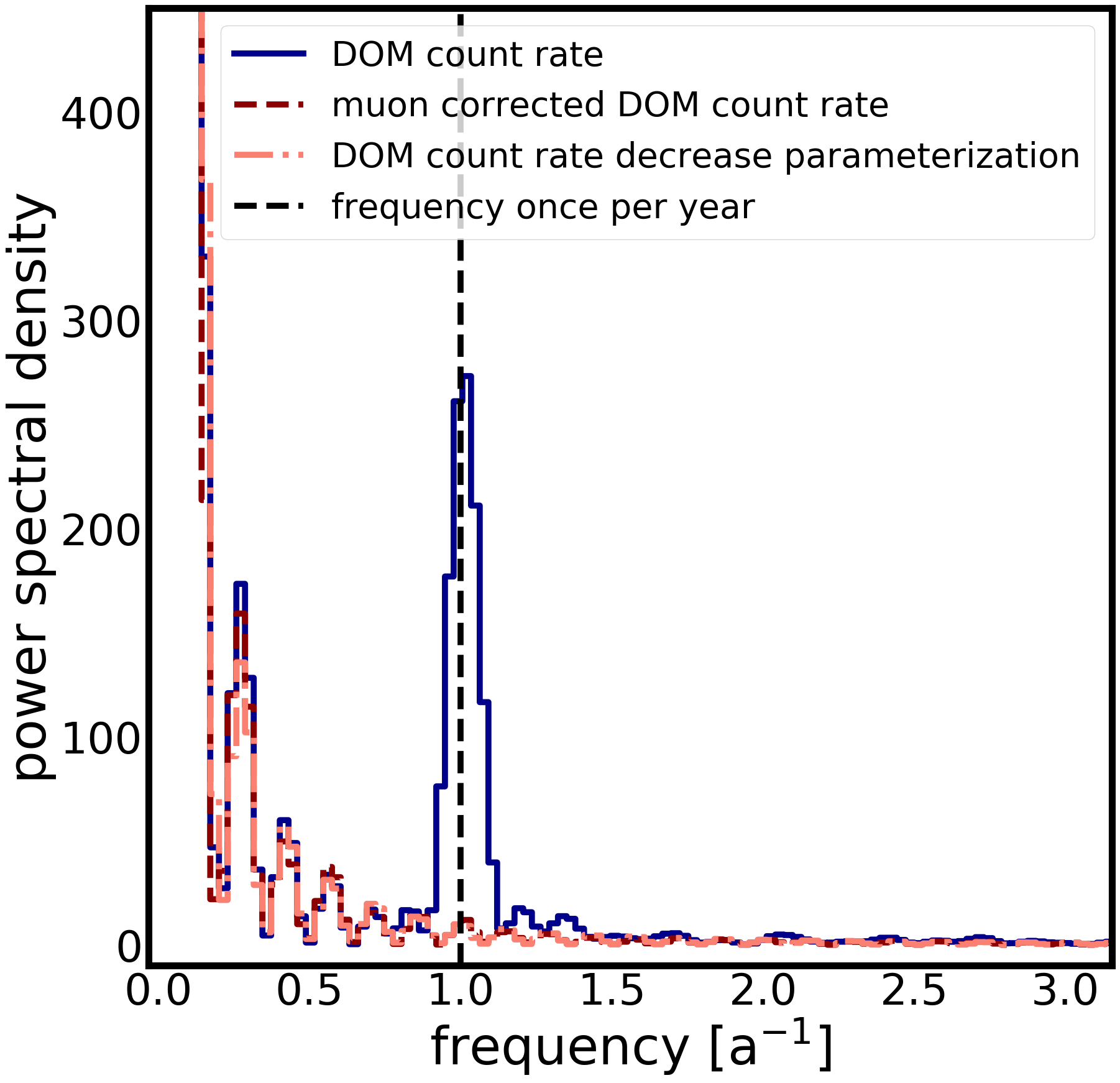}
\caption{\small Left: Summed rate in the completed IceCube detector with and without atmospheric muon suppression. Right: Lomb-Scargle periodogram for low frequency range (factor 5 oversampling). The frequency of 1/year (dashed line) sticks out.}
\label{fig:dataIC86}
\end{figure}

The Lomb-Scargle periodogram~\cite{bib:lomb,
bib:scargle} is designed to detect periodic signals in unevenly covered observations. The results are equivalent to fitting a sinusoidal function in each frequency bin. A white noise distribution at a fixed frequency is $\chi^2$ distributed under the null hypothesis. Otherwise, there are no obvious advantages w.r.t. a discrete Fourier transform~\cite{bib:vanderplaas}. 
Fig.~\ref{fig:dataIC86} (left) shows the rate per day for data taken with the completed IceCube detector  between 2013-2020. The data taking availability was 99.2\%. When rejecting problematic data taking periods, the uptime decreased to  96.6\%; the cosmic muon rejection further lowers the uptime to 94.6\%. Fig.~\ref{fig:dataIC86} (right) shows the same data as a power density in frequency space. Note that the rejection of atmospheric muons leads to a strongly decreased seasonal effect due to atmospheric pressure changes. The monotonically falling rate contribution mostly stems from a very slow decrease in photomultiplier noise rate with time.  This depth-dependent effect, which is leading to a diffractive pattern in the frequency space, may have to do with a decrease of triboluminescence as the ice releases stress. 

\section{Summary and outlook}
With our best current knowledge of the ice properties, detector effects, and neutrino cross sections, we determine effective volumes that are consistent with our previous result~\cite{bib:AA}. Using rare coincidences between neighboring modules, one can determine the average energy of interacting neutrinos in the event. The search for hidden or obscured supernovae in 13 years of data taken in time and frequency space will be published soon. As discussed in~\cite{bib:icrc2019}, IceCube will detect 99\% of  all Galactic core collapse supernovae
with neutrino fluxes equal or higher than in the conservative 8.8$M_\odot$ Hüdepohl model at $>9\sigma$ significance. 
We will use the periodogram method defined above to analyze all frequencies and quantify any excess from the expected background, as well as artifacts from changing run conditions and from the rejection of small faulty chunks of data. 

\bibliographystyle{ICRC}
\bibliography{main}

\clearpage
\section*{Full Author List: IceCube Collaboration}




\scriptsize
\noindent
R. Abbasi$^{17}$,
M. Ackermann$^{59}$,
J. Adams$^{18}$,
J. A. Aguilar$^{12}$,
M. Ahlers$^{22}$,
M. Ahrens$^{50}$,
C. Alispach$^{28}$,
A. A. Alves Jr.$^{31}$,
N. M. Amin$^{42}$,
R. An$^{14}$
% ...
\\

\noindent
% ...
$^{12}$ Universit{\'e} Libre de Bruxelles, Science Faculty CP230, B-1050 Brussels, Belgium \\
% ...
$^{14}$ Department of Physics and Laboratory for Particle Physics and Cosmology, Harvard University, Cambridge, MA 02138, USA \\
% ...
$^{17}$ Department of Physics, Loyola University Chicago, Chicago, IL 60660, USA \\
$^{18}$ Dept. of Physics and Astronomy, University of Canterbury, Private Bag 4800, Christchurch, New Zealand \\
% ...
$^{22}$ Niels Bohr Institute, University of Copenhagen, DK-2100 Copenhagen, Denmark \\
% ...
$^{28}$ D{\'e}partement de physique nucl{\'e}aire et corpusculaire, Universit{\'e} de Gen{\`e}ve, CH-1211 Gen{\`e}ve, Switzerland \\
% ...
$^{31}$ Karlsruhe Institute of Technology, Institute for Astroparticle Physics, D-76021 Karlsruhe, Germany  \\
% ...
$^{42}$ Bartol Research Institute and Dept. of Physics and Astronomy, University of Delaware, Newark, DE 19716, USA \\
% ...
$^{50}$ Oskar Klein Centre and Dept. of Physics, Stockholm University, SE-10691 Stockholm, Sweden \\
% ...
$^{59}$ DESY, D-15738 Zeuthen, Germany \\
% ...

\subsection*{Acknowledgements}

\noindent
USA {\textendash} U.S. National Science Foundation-Office of Polar Programs,
U.S. National Science Foundation-Physics Division,
U.S. National Science Foundation-EPSCoR,
Wisconsin Alumni Research Foundation,
Center for High Throughput Computing (CHTC) at the University of Wisconsin{\textendash}Madison,
Open Science Grid (OSG),
Extreme Science and Engineering Discovery Environment (XSEDE),
Frontera computing project at the Texas Advanced Computing Center,
U.S. Department of Energy-National Energy Research Scientific Computing Center,
Particle astrophysics research computing center at the University of Maryland,
Institute for Cyber-Enabled Research at Michigan State University,
and Astroparticle physics computational facility at Marquette University;
Belgium {\textendash} Funds for Scientific Research (FRS-FNRS and FWO),
FWO Odysseus and Big Science programmes,
and Belgian Federal Science Policy Office (Belspo);
Germany {\textendash} Bundesministerium f{\"u}r Bildung und Forschung (BMBF),
Deutsche Forschungsgemeinschaft (DFG),
Helmholtz Alliance for Astroparticle Physics (HAP),
Initiative and Networking Fund of the Helmholtz Association,
Deutsches Elektronen Synchrotron (DESY),
and High Performance Computing cluster of the RWTH Aachen;
Sweden {\textendash} Swedish Research Council,
Swedish Polar Research Secretariat,
Swedish National Infrastructure for Computing (SNIC),
and Knut and Alice Wallenberg Foundation;
Australia {\textendash} Australian Research Council;
Canada {\textendash} Natural Sciences and Engineering Research Council of Canada,
Calcul Qu{\'e}bec, Compute Ontario, Canada Foundation for Innovation, WestGrid, and Compute Canada;
Denmark {\textendash} Villum Fonden and Carlsberg Foundation;
New Zealand {\textendash} Marsden Fund;
Japan {\textendash} Japan Society for Promotion of Science (JSPS)
and Institute for Global Prominent Research (IGPR) of Chiba University;
Korea {\textendash} National Research Foundation of Korea (NRF);
Switzerland {\textendash} Swiss National Science Foundation (SNSF);
United Kingdom {\textendash} Department of Physics, University of Oxford.

\end{document}